April 2, 1992     ITP-SB-92-13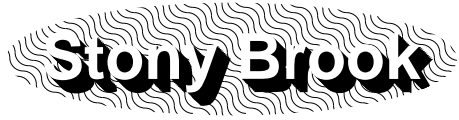

# THE N=4 STRING IS THE
# SAME AS THE N=2 STRING

W. Siegel[1]

*Institute for Theoretical Physics*
*State University of New York, Stony Brook, NY 11794-3840*

**ABSTRACT**

We redo the quantization of the N=4 string, taking into account the reducibility of the constraints. The result is equivalent to the N=2 string, with critical dimension D=4 and signature $(++--)$. The N=4 formulation has several advantages: the $\sigma$-model field equations are implied classically, rather than by quantum/$\beta$-function calculations; self-duality/chirality is one of the super-Virasoro constraints; SO(2,2) covariance is manifest. This reveals that the theory includes fermions, and is apparently spacetime supersymmetric.

---

[1] Work supported by National Science Foundation grant PHY 90-08936.
  Internet address: siegel@dirac.physics.sunysb.edu.

## 1. Summary

Recently there has been renewed interest in the N=2 string [1] as a description of self-dual Yang-Mills and gravity. The latter theories can be found by S-matrix calculations [2], or as $\sigma$-models (on the string) whose equations of motion (self-duality conditions) follow from requiring the vanishing of the $\beta$-function in quantum calculations [3]. A major drawback of this formalism is that the SO(2,2) covariance of the spacetime is obscure, since the N=2 world-sheet supersymmetry is not SO(2,2) covariant. The one state in such self-dual theories is represented by a "scalar" [4] which is actually the one surviving component of a tensor on-shell in an appropriate gauge. (For self-dual Yang-Mills, this is the scalar of Yang, and for self-dual gravity it's the Kähler scalar. In SO(2,2)=Sp(2)⊗Sp(2) spacetime, irreducible representations of the Poincaré group are self-dual or chiral, and have only one real physical component on shell.) For example, the Yang-Mills field can be expressed in terms of a Hertz potential as $A_a = \partial^b H_{ab} + \mathcal{O}(H^2)$. If the Hertz potential is constrained to be anti-self-dual, then the self-duality equation on the field strength becomes $\Box H + \mathcal{O}(H^2) = 0$. Then the on-shell gauge invariance $\delta H_{ab} = (\partial_{[a}\lambda_{b]} - dual) + \mathcal{O}(H)$ can be used to gauge away 2 of the 3 components of $H$. (Similar remarks apply for self-dual gravity, where the Kähler scalar is one component of a tensor in the same Lorentz representation as the anti-self-dual part of the Weyl tensor.) Extra ghosts are required to compensate for the fact that the usual gauge fields are derivatives of this "scalar." Since this component has nonvanishing helicity, it transforms nontrivially under Lorentz transformations even on shell, but these transformations have not been analyzed in the string theory. (For self-dual/chiral representations of SO(2,2), the little group is scale transformations GL(1), instead of compact U(1).)

The N=4 string [5] has drawn little attention because its critical dimension was thought to be negative. However, the reducibility of the constraints was unnoticed: In fact, if one first solves the N=2 subset of the N=4 constraints, the remaining constraints are redundant. (Thus, for example, arguments for the critical dimension based on the conformal anomaly will fail if they neglect the higher-generation ghosts resulting from this reducibility.) This should not be surprising, since the N=2 constraints in the critical dimension eliminate all excitations, reducing the string to a particle. As a result, the N=2 string can be treated in the N=4 formalism. The N=4 constraints, unlike the N=2 ones, are SO(2,2) covariant. The conformal-weight 1 constraints form the affine Lie algebra for Sp(2); this Sp(2) is half of the SO(2,2) spin, and its vanishing is exactly the constraint of self-duality/chirality. N=4 $\sigma$-models are finite [6-8]; as a result, the self-duality equations for Yang-Mills and gravity are found already at the classical level, required by the N=4 supersymmetry [6,7]. The explicit SO(2,2) also reveals a difference between Neveu-Schwarz and Ramond sectors: For example, for the open string, just as for N=1 (except for the self-duality), the Neveu-Schwarz sector describes self-dual Yang-Mills while the Ramond sector describes a Weyl Majorana spinor, so the whole string describes self-dual N=1 super Yang-Mills. (In the N=2 formalism the SO(2,2) transformations of these 1-component objects are obscure, while the $\sigma$-model approach is incapable of showing the existence of spinor fields.)

The N=4 string shares several properties with the Green-Schwarz formalism: (1) The critical dimension appears classically, since for D>4 there is no SO(D) symmetry. In Green-Schwarz the condition D=3,4,6,10 is found classically, which is not as strong as the quantum condition D=10, but more than for classical N=0 or 1 spinning strings, which say nothing, or the N=2 string, where even for D=4 the SO(2,2) symmetry is hidden. (2) $\sigma$-model field equations appear classically. In Green-Schwarz, some of the field equations appear classically, as a consequence of $\kappa$-symmetry, which is analogous to world-sheet supersymmetry. (3) The constraints are reducibile. However, they are not also second-class, as in Green-Schwarz. (4) GSO projection is unnecessary (there is no tachyon, and the Sp(2) constraint already makes the spinor chiral).

## 2. Constraints

The N=4 constraints can be written as

$$\mathcal{A} \equiv \tfrac{1}{2}P^{\alpha\beta'}P_{\alpha\beta'} - i\tfrac{1}{2}\Gamma^{\alpha\beta''}\Gamma'_{\alpha\beta''} = 0, \quad \mathcal{B}_{\alpha'\beta''} \equiv P^\gamma{}_{\alpha'}\Gamma_{\gamma\beta''} = 0, \quad \mathcal{C}_{\alpha''\beta''} \equiv \Gamma^\gamma{}_{(\alpha''}\Gamma_{\gamma\beta'')} = 0$$

(We work with holomorphic variables $P \equiv \partial X$ and $\Gamma$, suitable for describing the open string. As usual, the closed string requires a second set of such variables for the other handedness, while the heterotic string requires bosonic variables of the other handedness.) The N=2 subset of these constraints is given by $\mathcal{A}$ (Virasoro), $\mathcal{B}_{+'-''}$, $\mathcal{B}_{-'+''}$, and $\mathcal{C}_{+''-''}$ (GL(1)). (Note that the appropriate reality conditions for SO(2,2) imply that the N=4 gauged group is Sp(2)=SL(2)=SU(1,1), not SU(2), while the subgroup for N=2 is real scale transformations GL(1), not U(1).) Since we consider both the N=2 and N=4 string in D=4, the same set of physical variables ($P$ and $\Gamma$, 4 of each) is used.

To analyze the constraints, we can consider the usual Gupta-Bleuler-like analysis in terms of the physical oscillators, imposing the non-negative modes of the constraints on the states, and using the negative modes as generators of gauge transformations. In the N=2 theory, this analysis (as do other analyses) implies that the physical states are just the ground states, constructed without the oscillator modes of $P$ and $\Gamma$, and they satisfy the massless Klein-Gordon equation $p^2 = 0$. (The BRST proof [9] is equivalent to this way of phrasing Gupta-Bleuler. In the Ramond sector, there are also fermionic constraints in terms of the zero-modes $\gamma$ and $p$.)

In the N=4 theory, we can make the same type of analysis, but consider the N=2 subset of the constraints first, where the analysis is the same, since the physical variables $\Gamma$ and $P$ are the same, and this subset has the same representation in terms of them as in the N=2 theory. Since the physical states are again restricted to ground states, the positive modes of the remaining N=4 constraints $\mathcal{B}_{+'+''}$, $\mathcal{B}_{-'-''}$, $\mathcal{C}_{+''+''}$, and $\mathcal{C}_{-''-''}$ then automatically vanish on the physical states, so these constraints are redundant, and the negative modes generate no gauge transformations of the ground states. In the Neveu-Schwarz sector, having eliminated all the non-zero modes of $P$ and all the modes of $\Gamma$ (which is half-integrally moded), we are left with the constraint $p^2 = 0$.

In the Ramond sector, the N=2 constraints again eliminate all non-zero modes of $P$ and $\Gamma$, making all non-zero modes of the remaining N=4 constraints redundant. This leaves only the zero modes of the N=4 constraints, to be imposed on states which are a representation of just the zero modes of $P$ and $\Gamma$, i.e. a Dirac spinor: In Sp(2)⊗Sp(2) notation, the representation of $\gamma_{\alpha\alpha''}$ is $|\Lambda\rangle = \lambda^\alpha|\alpha\rangle + \lambda^{\alpha''}|\alpha''\rangle$, where $\gamma_{\alpha\alpha''}|\beta\rangle = C_{\alpha\beta}|\alpha''\rangle$ and $\gamma_{\alpha\alpha''}|\beta''\rangle = C_{\alpha''\beta''}|\alpha\rangle$. (The $C$'s are the antisymmetric, hermitian metrics of the Sp(2)'s.) We are thus left with an N=4 spinning particle. The zero mode of $\mathcal{A}$ is again the massless Klein-Gordon equation. The zero mode of $\mathcal{C}$ is one of the two Sp(2)'s of spin, and it kills one of the two Weyl spinors ($\lambda^{\alpha''}$) making up the Dirac spinor. Finally, the zero mode of $\mathcal{B}$ imposes the massless Dirac equation on the surviving Weyl spinor (and sets to zero the gradient of the other Weyl spinor, which was already eliminated by the $\mathcal{C}$ constraint). If we had instead solved just the N=2 constraints, we would have been left with an N=2 spinning particle (as defined by the zero modes of the N=2 string), and the results would be the same (though not in covariant form) with an appropriate choice of the normal-ordering constant in $\mathcal{C}_{+''-''}$.

In both sectors we are left with one physical polarization: In the Neveu-Schwarz sector we have what appears to be a scalar but is actually the one surviving component on-shell of self-dual Yang-Mills; in the Ramond sector we have the one on-shell component of a Weyl Majorana spinor. In the N=4 formalism the Lorentz covariance of the spinor is manifest because, as for the N=1 string, the ground state of the Ramond sector is itself the spinor. On the other hand, in this analysis of the Neveu-Schwarz sector in terms of just the physical operators $P$ and $\Gamma$, the Lorentz transformations of the vector are just as obscure in the N=4 formalism as in the N=2. This is related to the fact that, as for the N=0 and N=1 strings, the vector is not the naive ground state, but is created by acting with oscillators on a "tachyon." This problem will be resolved when the Lorentz covariant ghosts are understood, since the true ground state in the Neveu-Schwarz sector is the zero-momentum Yang-Mills ghost, which is in the BRST cohomology at unphysical ghost number. In the N=2 formalism the SO(2,2) transformations of the states are obscure, so the one state of the Ramond sector is confused with the one state of the Neveu-Schwarz sector. (Without Lorentz transformations, statistics is also obscure.) This "equivalence" is usually explained by the ambiguity in redefining $\mathcal{A}$ by adding an arbitrary constant times the derivative of $\mathcal{C}_{+''-''}$; but this redefinition is forbidden in the N=4 formulation by Sp(2) covariance.

Thus, at least at the free level, we have a spacetime-supersymmetric theory, self-dual super Yang-Mills, consisting of self-dual Yang-Mills and a real chiral spinor (for the open string; similar remarks for self-dual supergravity apply for the closed/heterotic case, whose spectrum can be obtained by direct products of open string spectra). The description is the same as for SO(3,1), except that the two types of Weyl spinor (index) are now both real, instead of being complex conjugates of each other. Thus, anti-chiral superfield strengths can consistently be set to vanish while keeping the corresponding chiral superfield. So, an N=1 spacetime supersymmetric multiplet consists of real component field strengths $A_{(\alpha_1...\alpha_n)}$ and $B_{(\alpha_1...\alpha_{n+1})}$, totally symmetric in their indices, and killed by $p^{\alpha\alpha'}$ contracted with any index, so each has one physical polarization. The chirality of the fermions leads to the same kind of problem writing actions as self-duality for the bosons: The usual action $\lambda^\alpha p_{\alpha\alpha'} \lambda^{\alpha'}$ would require both chiralities. There is a corresponding difficulty in defining norms, but this should not be surprising in a theory with 2 time dimensions. (Notice that norms are not necessary in finding the physical states in the way we have defined the Gupta-Bleuler procedure; analogously, the definition of BRST cohomology does not make use of the norm.) Since super Yang-Mills in D=4 is also spacetime conformal, we have the interesting result that a world-sheet superconformal theory results in a spacetime superconformal theory. For SO(2,2), the superconformal group is SL(4|N) (for N supersymmetries; SL(4)=SO(3,3)); a manifestly SL(4|N) formulation of these self-dual

theories might be useful. Supersymmetry at the interacting level is expected to follow from the type of construction used for N=1 strings [10]; however, a Lorentz covariant supersymmetry generator will require the N=4 formalism and its ghosts.

Although the above demonstration of the redundancy of the N=4 constraints is sufficient to show equivalence to N=2, an equivalent covariant analysis is needed for a covariant treatment of ghosts and BRST. The reducibility of the constraints (at least at the first level) is most easily formulated with harmonic superspace [8] (although we use a slightly different harmonic superspace than in [8]). In addition to the usual world-sheet coordinates $z$ and $\theta^{\alpha'\alpha''}$, we introduce $u^{\alpha''}$. Then the usual supersymmetric derivative $d_{\alpha'\alpha''}$ and superfield $\Psi_{\alpha\alpha''} = \Gamma_{\alpha\alpha''} + \theta^{\alpha'}{}_{\alpha''}P_{\alpha\alpha'} + z\text{-}derivative\ terms$ are replaced with

$$\Psi_\alpha \equiv u^{\alpha''}\Psi_{\alpha\alpha''}, \quad d_{\alpha'} \equiv u^{\alpha''}d_{\alpha'\alpha''}, \quad d_{\alpha'}\Psi_\alpha = 0$$

The constraints are then

$$\mathcal{T} \equiv \tfrac{1}{2}\Psi^\alpha\Psi_\alpha \equiv u^{\alpha''}u^{\beta''}\mathcal{T}_{\alpha''\beta''} = 0$$

where the $\theta$ expansion of $\mathcal{T}_{\alpha''\beta''}$ gives $\mathcal{C}$, $\mathcal{B}$, $\mathcal{A}$. The reducibility condition

$$\Psi_\alpha T = 0$$

then follows from the commutation relations

$$\{\Psi_\alpha(1), \Psi_\beta(2)\} = u_1^{\alpha''}u_{2\alpha''}d^4\delta^5(z,\theta)$$

which vanish at $u_1 = u_2$. There is an infinite chain of such reducibilities (from multiplying repeatedly by $\Psi_\alpha$ at each ghost level, since $\Psi_\alpha\Psi_\beta \sim C_{\alpha\beta}T$), as well as new reducibility conditions which show up at the second ghost level. The reducibility $\Psi_\alpha T = 0$ includes such conditions as $P^{\alpha\alpha'}\mathcal{B}_{\alpha'\alpha''} + \cdots = 0$, which is of the same form as the infinite chain of reducibility conditions which appear for the analogous fermionic constraint ($\not{P}D$) in the Green-Schwarz string. This suggests that an understanding of the covariant quantization of the N=4 string might help with the Green-Schwarz case.

## ACKNOWLEDGMENTS

I thank Nathan Berkovits, Jim Gates, and Martin Roček for directing me to the N=2 string, educating me about various points regarding it and N=4 $\sigma$-models, and for many discussions.